\documentclass[12pt]{article}

\usepackage{graphicx}
\usepackage{amsmath,amssymb,amsfonts}
\usepackage{hyperref}
\usepackage{lmodern}
\usepackage{amssymb}
\usepackage{booktabs}

\renewcommand{\theequation}
{\arabic{section}.\arabic{equation}}

\makeatletter
\def\eqnarray{ \stepcounter{equation} \let\@currentlabel=\theequation
 \global\@eqnswtrue
 \global\@eqcnt\z@
 \tabskip\@centering
 \let\\=\@eqncr
 $$\halign to \displaywidth\bgroup\@eqnsel\hskip\@centering
 $\displaystyle\tabskip\z@{##}$&\global\@eqcnt\@ne
 \hfil$\displaystyle{{}##{}}$\hfil
 &\global\@eqcnt\tw@$\displaystyle\tabskip\z@{##}$\hfil
 \tabskip\@centering&\llap{##}\tabskip\z@\cr}
\makeatother

\makeatletter
\def\@arrayacol{\edef\@preamble{\@preamble \hskip .5\arraycolsep}}
\def\array{\let\@acol\@arrayacol \let\@classz\@arrayclassz
\let\@classiv\@arrayclassiv \let\\\@arraycr\def\@halignto{}\@tabarray}
\makeatother

\makeatletter
\newcounter{subeqncnt}
\def\thesubeqncnt{\alph{subeqncnt}}
\def\subequations{\begingroup%
   \stepcounter{equation}\edef\@tempa{\theequation}%
   \let\c@equation\c@subeqncnt\c@subeqncnt\z@
   \edef\theequation{\@tempa\noexpand\thesubeqncnt}}

\makeatother

\newcommand{\be}{\begin{equation}}
\newcommand{\ee}{\end{equation}}

\newcommand{\beqa}{\begin{eqnarray}}
\newcommand{\eeqa}{\end{eqnarray}}
\newcommand{\nn}{\nonumber}

\def\CD {{\cal D}}

\def\CO {{\cal O}}

\begin{document}

\setlength{\baselineskip}{7mm}

\hfill  NRCPS-HE/2017-52

\vspace{1cm}
\begin{center}
{\Large ~\\{\it  Gravity  with a Linear Action\\

\vspace{1cm}
and \\
\vspace{1cm}
Gravitational Singularities 

}

}

\vspace{2cm}

{\sl   George Savvidy

\bigskip
  {\sl Institute of Nuclear and Particle Physics\\
  Demokritos National Research Center\\
 Ag. Paraskevi,  Athens, Greece}

\bigskip

}
\end{center}
\vspace{1cm}

\centerline{{\bf Abstract}}

\vspace{0.5cm}

\noindent
Motivated by  quantum mechanical considerations we earlier suggested  an alternative action for discretised quantum gravity which has a dimension of length.  It is the so called "linear" action. The proposed action is a "square root" of the classical area action in gravity and has in front of the action a new constant of dimension one. Here we shall consider the continuous limit of the discretised linear action.   We shall demonstrate that in  the modified theory of gravity there appear space-time regions of the Schwarzschild radius scale  which are unreachable by test particles. These regions are located in the places where standard theory of gravity has singularities.  We are confronted  here with a drastically new concept that there may exist space-time regions which are excluded from the physical scene, being physically unreachable by test particles or observables. If this concept is accepted, then it seems plausible that the gravitational singularities are  excluded from the modified theory.

\newpage

\pagestyle{plain}


\section{\it Introduction}

Unification of gravity with other fundamental forces within the superstring theory stimulated the interest to the theory of quantum gravity and to physics at Planck scale 
\cite{sacharov,Buchdahl,Starobinsky:1980te,Adler:1982ri,Gasperini:1992em}. In particular, string theory predicts modification of the gravitational action 
at Planck scale with additional high derivative terms. This allows to ask fundamental questions concerning physics at Planck scale referring to these effective actions  and, in particular, one can try to understand how they influence the gravitational singularities 
\cite{Penrose:1964wq,Christodoulou,Hawking1,Hawking2,Hawking3,HawkingPenrose,Robertson,Raychaudhure,Komar,Markov:1982ed,anini,Brandenberger:1988aj,Alvarez:1984ee,Baierlein:1962zz,Brandenberger:1993ef,Deser:1998rj}.
 
It is appealing to extend  this approach to different modifications of classical gravity which follow from the string theory and also to develop an alternative approach which is based on new geometrical principles \cite{Savvidy:1995mr,Ambjorn:1996kk,Savvidy:1997qf}\footnote{See also the references \cite{Ambjorn:1985az,Ambartsumian:1992pz, Ambjorn:1997ub,Savvidy:2015ina}.}. These geometrical principles allow to extend the notion of the Feynman integral over paths to an integral over space-time manifolds with the property that in cases when a manifold degenerates into a single world line the quantum mechanical amplitude becomes proportional to the length of the world line. This approach to quantum gravity is based on the idea that quantum mechanical amplitudes should be proportional to the "length" of the quantum fluctuations when they are of geometrical origin.
In other words, in this limit the action should reduce to the action of the relativistic particle \cite{Savvidy:2015ina}.  

The discretised  version of the linear action can be derived from  geometrical principles: 
{{\bf $\alpha )$}} the coincidence of the transition amplitude 
with the Feynman path amplitude for a manifold which is degenerated to a single world line and {{\bf $\beta )$}} the continuity of the transition 
amplitudes under the manifold deformations \cite{Savvidy:1995mr,Ambjorn:1996kk,Savvidy:1997qf}. In accordance with $\alpha )$ the quantum mechanical amplitude should be proportional to the {\it length of the space-time manifold} and therefore  it must be proportional to the linear combination of the lengths of all edges of 
discretised space-time  manifold
$
 \sum_{<i,j>} \lambda_{ij}\cdot \Theta_{ij}
$, 
where $\lambda_{ij}$ is the length of the edge between 
two vertices $<i>$ and $<j>$, summation is over all
edges $<i,j>$  and $\Theta_{ij}$ is unknown angular factor, which can be defined through the continuity principle $\beta )$.  The deficit angel $\Theta_{ij}$ should  vanish in cases when the triangulation has  $flat$ surfaces, thus 
$\sum_{<i,j>}\lambda_{ij}\cdot \sum (2\pi - 
\sum \beta_{ijk}),$
where  $\beta_{ijk}$ are the angles on the cone which appear in the normal section of the edge $<ij>$ and triangle $<ijk>$. Combining terms belonging to a given triangle
$<ijk>$ we shall get a sum $\lambda_{ij}+\lambda_{jk}+\lambda_{ki}=
\lambda_{ijk}$ which is equal to the perimeter of the triangle $<ijk>$ 
\cite{Savvidy:1995mr,Ambjorn:1996kk,Savvidy:1997qf}:
\begin{equation}
S_L = \sum_{<ijk>}\lambda_{ijk} \cdot \omega^{(2)}_{ijk}
\label{lamb1}
\end{equation}
and where $\omega^{(2)}_{ijk}$ is the deficit angle associated with
the triangle $<ijk>$. 

The linear character of the  action requires the 
existence of a new fundamental coupling constant $m_P$ of dimension $ 1/cm$.
It is natural to call these amplitudes "linear" or "gonihedric" because the definition contains the sum of products of the characteristic lengths and deficit angles.  
Thus the principles of {\it linearity} and  {\it continuity}  allow to define the linear action  which can be considered as a "square root" of classical Regge area action in gravity
\cite{Regge:1961px,wheeler}. 
It is useful to compare the linear action (\ref{lamb1}) with the Regge area action \cite{Regge:1961px,wheeler}.  
In (\ref{lamb1}) the {\it perimeter}  of a triangle is  multiplied by the
deficit angle.   In the Regge action  the {\it area} of the  
triangle $\sigma_{ijk}$ is  multiplied by the deficit angle
\begin{equation}
S_R= \sum_{<ijk>} \sigma_{ijk} \cdot \omega^{(2)}_{ijk}
\label{area}
\end{equation}
and represents the discretised version of the standard continuous area action in gravity: 
\be\label{standgravity}
S_A =-{  c^3\over 16 \pi G } \int R \sqrt{-g} d^4x ,
\ee
where 
$
G=  6.67 \cdot 10^{-8} {cm^3 \over g~ sec^2}
$
is the gravitational coupling constant.  The dimension of the measure $ [ \sqrt{-g} d^4x]$ is 
$cm^4$, the dimension of the scalar curvature  $[R] $ is ${1/ cm^2} $, thus the 
integral $\int R \sqrt{-g} d^4x$ has dimension $cm^2$ and measures  the "area" of the universe.
The coupling constant ${  c^3\over 16 \pi G }$ in front of the integral has therefore the dimension $  {g  \over sec} $.  

It is unknown to the author how to derive a continuous limit  of the linear action (\ref{lamb1}) in a unique way. In this circumstance we shall try to 
construct a possible linear action for a smooth space-time universe by using the available geometrical invariants.  Any expression 
which is quadratic in the curvature tensor and includes two derivatives can be a candidate for the linear action.  These invariants  have the following form:
\beqa\label{lineargravity2}
 I_1= -{1\over 180}R_{\mu\nu\lambda\rho;\sigma} R^{\mu\nu\lambda\rho;\sigma} ,~~~~~~~~~~I_2=+{1\over 36} R_{\mu\nu\lambda\rho}  \Box  R^{\mu\nu\lambda\rho}~,  
\eeqa
and we shall consider a linear combination of the above expressions\footnote{The general form of the action is presented in the Appendix.}:
 \beqa\label{lineargravity}
 && S_L=  -M c
  \int {3  \over 8 \pi} (1-\gamma) \sqrt{I_1 +\gamma I_2 }~\sqrt{-g }  d^4x,
\eeqa
where we introduced the corresponding mass parameter $M$ and the dimensionless parameter $\gamma$. The dimension of the  invariant  $[\sqrt{I_1 + \gamma I_2 }] $ is ${1/ cm^3} $, thus the integral $\int \sqrt{I_1 + \gamma I_2 } \sqrt{-g} d^4x$ has the dimensions of $cm$ and measures  the "length" of the universe. The  expression (\ref{lineargravity}) fulfils our basic physical requirement on the action that it should have the dimension of length and should be similar to the action of the relativistic particle 
\be\label{relatparticle}
S=-M c \int  d s= -M c^2 \int \sqrt{1-{\vec{v}^2\over c^2 }}~ dt.
\ee
Both expressions contain the geometrical invariants which are not in general positive definite under the square root operation. In the relativistic particle case (\ref{relatparticle}) the expression under the root becomes negative for a particle moving with a velocity which exceeds the velocity of light. In that case the action develops an imaginary part and quantum mechanical suppression of amplitudes prevents a particle from exceeding the velocity of light \cite{Pauli:1941zz,Feynman:1949hz,Feynman:1949zx}. 
A similar mechanism  was implemented in the Born-Infeld modification of electrodynamics with the aim to prevent the appearance of infinite electric fields \cite{borninfeld}.
 
 One can expect that 
in case of modified gravity with linear action (\ref{lineargravity}) there may appear space-time regions which are unreachable by the test particles
as in that regions the expression under the root were negative. If these "locked"  space-time regions happen  to appear and if that space-time regions 
include singularities, then one can expect  
that the gravitational singularities are naturally excluded from the theory  due to the fundamental principles of quantum mechanics. The question of consistency  of the new action principle, if it is  the 
right one, can only be decided by their physical consequences. 

In the next sections we shall consider the  black hole singularities and the physical effects which are induced  by the inclusion of the linear action (\ref{lineargravity}). As
we shall see, the expression under the root becomes negative in the region which is smaller than the Schwarzschild radius $r_g$ and includes the singularities. For the observer which is far away from the horizon the linear action perturbation induces the advance precession of the perihelion and has a profound influence on the physics near the horizon. We are confronted here with a drastically unusual concept that there may exist 
space-time regions which are excluded from the physical scene, being physically unreachable by test particles or observables. If one accepts 
this concept, then it seems plausible that the gravitational singularities are  excluded from the modified theory. In this paper we have only taken the first steps to describe the phenomena  which are caused by the additional linear term in the gravitational action proposed in \cite{Savvidy:1995mr,Ambjorn:1996kk,Savvidy:1997qf}.

  \section{\it Schwarzschild Black Hole Singularities}

\begin{figure}
\begin{center}
\includegraphics[width=10cm]{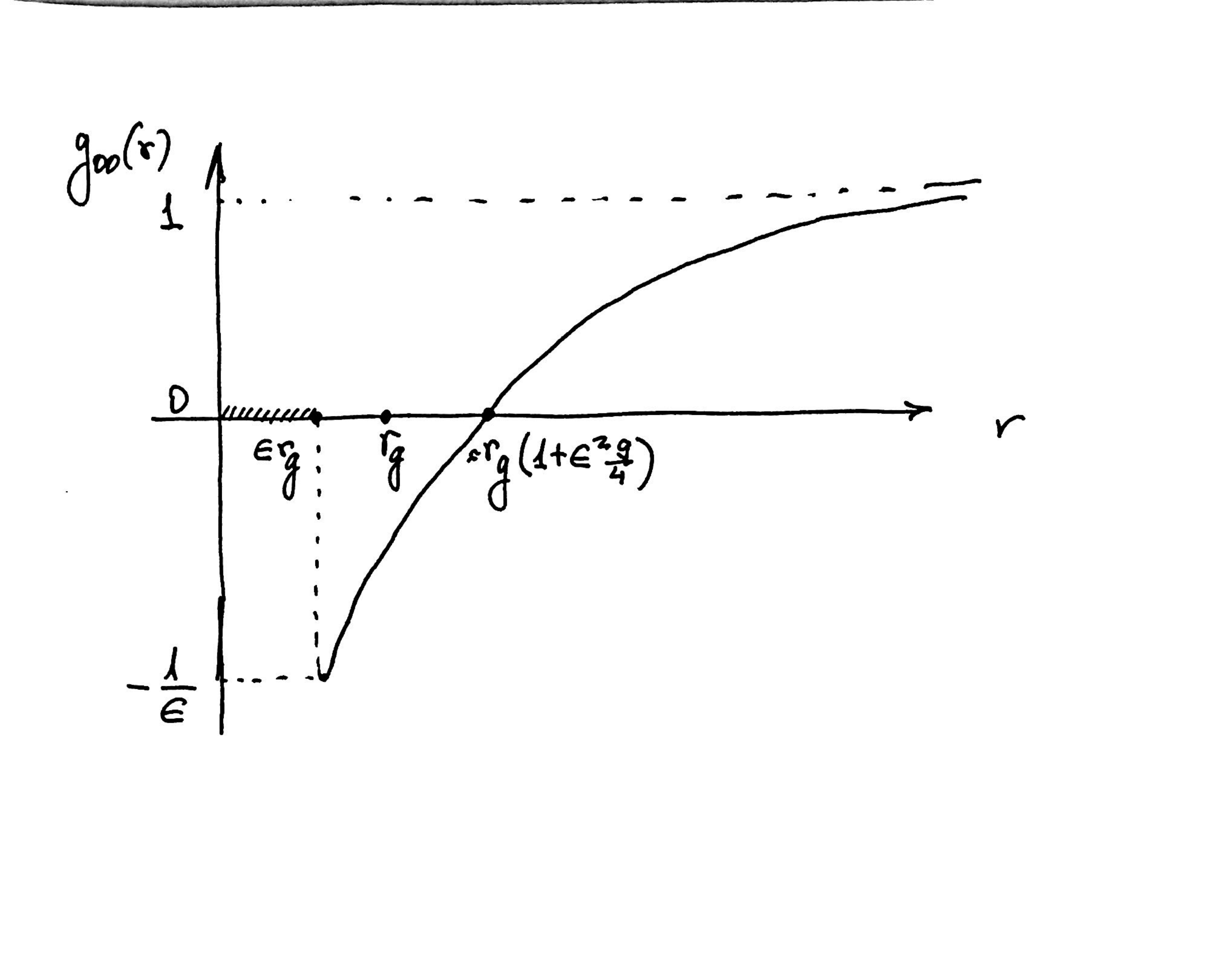}~~~~~~~~~~
\caption{
The graphic of the potential function $g_{00}=1-{r_g\over r} -  \Big[1-
\Big(1 -\varepsilon  {r_g\over r} \Big)^{3/2} \Big]^2 $. At $r \rightarrow \infty$ the 
$g_{00} \rightarrow 1$ and at $r \rightarrow \varepsilon r_g $ the 
$g_{00} \rightarrow -{1\over \varepsilon}  $.  
}
\label{fig1}
\end{center}
\end{figure}

The modified action which we shall consider is a sum
\beqa\label{actionlineargravity23}
 && S =-{  c^3\over 16 \pi G  } \int R \sqrt{-g} d^4x 
-  M c  \int {3  \over 8 \pi}(1-\gamma)\sqrt{ I_1+ \gamma I_2
}~\sqrt{-g }  d^4x,
\eeqa
where we introduced a dimensionless parameter $\gamma$ in order to consider a general linear combination of the invariants. 
The additional linear term has high derivatives of the metric and the equations of motion which follow from the variation of this action are much more complicated than in the standard case, and we shall consider them in the forthcoming sections. Here we shall consider the perturbation of the Schwarzschild solution which is induced  by the the additional term in the action and try to understand how it influences the black hole physics and the singularities. The Schwarzschild solution has the form 
\be\label{schwarz}
ds^2 = (1-{r_g\over r}) c^2 dt^2 - (1-{r_g\over r})^{-1} dr^2 -r^2 d\Omega^2~,
\ee
where 
$
g_{00} =1-{r_g\over r},~g_{11} =-(1-{r_g\over r})^{-1},~g_{22} =-r^2,~g_{33} =-r^2 \sin^2\theta,
$
and 
$$
r_g = {2GM \over c^2},~~~~ \sqrt{-g} = r^2 \sin\theta.
$$
The nontrivial quadratic curvature invariant in this case has the form
\beqa
&&I_0={1\over 12} R_{\mu\nu\lambda\rho}  R^{\mu\nu\lambda\rho} = ({r_g \over r^3})^2  \eeqa
 and shows that the singularity located at $r=0$ is actually a curvature singularity.
 The event horizon is located where the metric component $g_{rr}$ diverges, that is,  at 
$
r_{horizon}=  r_g . 
$
The expressions for the two curvature polynomials (\ref{lineargravity2}) of our interest are\footnote{It should be stressed that all other invariant polynomials of the same dimensionality can be expressed in terms of $I_1$ and $I_2$, and they are given in Appendix.}:
\beqa
I_1= {r^2_g( r -  r_g )  \over  r^9},~~~~~I_2={r^3_g \over  r^9}~, 
\eeqa
and on the Schwarzschild solution the action acquires additional term of the form 
\beqa\label{lineargravityS}
S =  -M c^2    \int  {3  \over 2} \varepsilon  \sqrt{  1-\varepsilon {  r_g   \over  r }}~ {r_g \over r^2} dr dt~,
\eeqa
where
\be
\varepsilon = 1-\gamma . \nn
\ee
As one can see, the expression under the square root in (\ref{lineargravityS}) becomes negative at
\be 
r < \varepsilon r_g, ~~~~~0 < \varepsilon \leq 1
\ee
and defines  the region which is unreachable by the test particles.  The size of the region  depends on the parameter $\varepsilon$ and is smaller than the gravitational radius $r_g$ (see Fig. \ref{fig1}) . 

This result seems to have profound consequences on the gravitational singularity at $r=0$. 
In a standard interpretation of the singularities, which appear in spherically symmetric gravitational collapse,  the singularity at $r=0$ is hidden in the sense that no signal from it can reach infinity. The singularities
are not visible for the outside observer, but hidden behind an event horizon. In that interpretation the singularities are still present in the theory. In the suggested 
scenario it seems possible to eliminate  the singularities from the theory based on the fundamental principles of quantum mechanics.   The singularities are excluded from the theory on the same level as the motion of particles with a velocity which
exceeds the speed of light. 

The quantum mechanical amplitude in terms of the path integral 
has the form 
\be
\Psi = \int e^{{i \over  \hbar } S[g]} \CD g_{\mu\nu}(x) \nn,
\ee 
where integration is over all diffeomorphism nonequivalent metrics.
For the Schwarzschild massive object which is at rest we can find the expression for the action
\be
S = - M c^2    \int^{\infty}_{\varepsilon r_g} {3  \over 2}\varepsilon  \sqrt{  1-\varepsilon {  r_g   \over  r }}~ {r_g \over r^2} dr dt \\
=M c^2   t   
\ee
and confirm that it is proportional to the length $t$  of the space-time trajectory, as it should be for the relativistic particle  at rest, 
so that the corresponding amplitude can be written in the form
\be
\Psi \approx  \exp{ ({i   \over   \hbar } \sum_n M_n c^2 t)},
\ee
where the summation is over all bodies in the universe. 

The perturbation (\ref{lineargravityS}) generates a contribution to the distance invariant $ds$ in (\ref{schwarz}) of the form
\be\label{hatper}
{3  \over 2}  \int^{\infty}_{r} \varepsilon \sqrt{  1-\varepsilon {  r_g   \over  r }}~ {r_g \over r^2} dr =    \Big[1-
\Big(1 -\varepsilon  {r_g\over r} \Big)^{3/2} \Big]
\ee
and allows to calculate  the correction to the purely temporal component of the metric tensor (\ref{schwarz})  caused by the additional term in the  action
\beqa\label{main}
&&g_{00} =1-{r_g\over r} -    \Big[1-
\Big(1 -\varepsilon  {r_g\over r} \Big)^{3/2} \Big]^2  .
\eeqa
The  equation used to determine gravitational time dilation near a massive body is modified in this case and the  proper time between events is defined now by the equation
\be
d \tau = \sqrt{g_{00}} dt = \sqrt{1-{r_g\over r} -    \Big[1-
\Big(1 -\varepsilon  {r_g\over r} \Big)^{3/2} \Big]^2} dt
\ee
and therefore
$
d \tau   \leq dt,
$ as in standard gravity. 
It follows from (\ref{main}) that near the gravitational radius $r \approx r_g$ a purely temporal component of the metric tensor has the form 
\beqa
g_{00} ~\approx ~1-{r_g\over r} - \varepsilon^2 {9 \over 4}   \Big( {r_g\over r} \Big)^{2} 
 +  \CO( \varepsilon^3)
\eeqa
and the infinite red shift which appears in the standard case at $r=r_g$ 
now appears at 
\be
r \approx  ~ r_g (1+ {9 \over 4} \varepsilon^2)  +\CO(\varepsilon^4) .
\ee 
To define the perturbation of the  trajectories of the test particles we shall study the behaviour  of the solutions of the Hamilton-Jacobi equation for geodesics, which is modified by the perturbation of the metric: 
\beqa
&g^{\mu\nu} {\partial A \over \partial x^{\mu}}{\partial A \over \partial x^{\nu}}~
  = g^{00}  \Big({\partial A \over c \partial t}\Big)^2-
{1\over g^{00}} \Big({\partial A \over  \partial r}\Big)^2 - {1\over r^2}\Big({\partial A \over  \partial \phi}\Big)^2 =m^2 c^2.\nn
\eeqa
The solution has the form
\be
A = -E t + L \phi +A(r),
\ee  
where  $E$ and $L $ are the energy and angular momentum of the test particle and  
\be\label{geodesic}
A(r)= \int \Big[ \Big( g^{00} {E^2 \over c^2} - m^2 c^2 -{L^2 \over r^2} \Big) g^{00} \Big]^{1/2} dr .
\ee
In the non-relativistic limit $E=E^{'} + m c^2, E^{'} \ll m c^2$, and in terms of a new coordinate 
$r(r-r_g)=r^{'}$
we shall get 
\be
A(r) \approx \int \Big[   ( {E^{'2} \over c^2}  +2 E^{'} m) 
+{1 \over r^{'}} (4 E^{'} m r_g + m^2 c^2 r_g) 
- {1 \over r^{' 2}} \Big(  L^2 - {3  \over 2} m^2 c^2 r^2_g (1 + {3  \over 2}  \varepsilon^2)    \Big) \Big]^{1/2} dr^{'} .
\ee
The geodesic trajectories are defied by the equation 
$  \phi + \partial  A(r) / \partial L = Const$  and the advance precession of the 
perihelion $\delta \phi$  expressed in radians per revolution is  given by the expression 
\be
\delta \phi =  {3 \pi m^2 c^2 r^2_g  \over 2 L^2 } (1 + {3  \over 2}  \varepsilon^2)=
{6 \pi G M   \over c^2 a (1-e^2) } (1 + {3  \over 2}  \varepsilon^2),
\ee
where $a$ is the semi-major axis and $e$ is the orbital eccentricity. As one can see from the above result, the precession is advanced by the additional factor $1 +{3  \over 2}  \varepsilon^2$. The upper bound on the value of  $\varepsilon$ can be extracted from the observational data for the advanced precession of the Mercury perihelion,  which is $42,98 \pm 0,04$ seconds of arc per century, thus
$$
\varepsilon \leq 0,16~.
$$
For the light propagation we shall take $m^2 =0$,~ $E= \omega_0$,~$L =  \rho~  \omega_0 / c$ in (\ref{geodesic}):
\be
A(r)=  {\omega_0 \over c} \int \sqrt{  ( g^{00}   - {\rho^2    \over   r^2} ) g^{00} }  dr  \approx   {\omega_0 \over c} \int \sqrt{ 1+ 2 {r_g\over r} - {\rho^2    \over   r^2}   }  dr +\CO( \varepsilon^2  r^2_g /r^2). 
 \ee
The trajectory is defined by the equation $ \phi + \partial  A(r) / \partial \rho = Const$ and in the given approximation the deflection of light  ray remains unchanged:
\be
\delta \phi =  2 {r_g\over \rho},
\ee
where $\rho$ is the distance from the centre of gravity. The deflection angle is not influenced by the perturbation, which is of order $\CO( \varepsilon^2  r^2_g / \rho^2)$,  and does not impose a sensible constraint on $\varepsilon$. 
In the next sections we shall consider perturbation of the Reissner-Nordstr\"om and the Kerr solutions.

\vspace{1cm}

 \section{\it Reissner-Nordstr\"om Solution}

The Reissner-Nordstr\"om solution has the form 
\be\label{reissnernordstrom}
ds^2 = (1-{r_g\over r} + {r^2_{Q}\over r^2}) c^2 dt^2 - (1-{r_g\over r} + {r^2_{Q}\over r^2})^{-1} dr^2 -r^2 d\Omega^2~,
\ee
where 
$$
r_g = {2GM \over c^2},~~~~r^2_{Q} = { Q^2 G \over c^4},~~~~~~\sqrt{-g} = r^2 \sin\theta.
$$
The nontrivial quadratic curvature invariant is
\beqa
&&I_0={1\over 12} R_{\mu\nu\lambda\rho}  R^{\mu\nu\lambda\rho} = {3 r^2 r^2_g  -12 r r_g r^2_Q +14 r^4_Q\over 3 r^8},  
\eeqa
 and it shows that the singularity is located at $r=0$.
The event horizon and internal Cauchy horizon are located where the metric component $g_{rr}$ diverges: 
\be
r_{\pm}= {1\over 2}(r_g \pm \sqrt{r_g^2 -4 r^2_Q}).
\ee
The solutions with $r_Q > r_g/2$ represent a naked singularity.  As we shall see below, at these charges $r_Q$ the linear action becomes a complex valued function  and  prevents the appearance of the naked singularities.

The expression for the two curvature polynomials of our interest are:\beqa
&&I_1= {  (r^2 - r r_g + r_Q^2) (45 r^2 r_g^2 - 216 r r_g r_Q^2 + 304 r_Q^4) \over 45r^{12}  },\nn\\
&&I_2={9 r^3 r_g^3 + 36 r^3 r_g r_Q^2 - 96 r^2 r_g^2 r_Q^2 - 88 r^2 r_Q^4 + 
 264 r r_g r_Q^4 - 200 r_Q^6 \over 9 r^{12}}. 
\eeqa
It is convenient to introduce the  dimensionless quantities:
\be\label{units}
\hat{r} = {r \over r_g},~~~~~\hat{r_Q}= {r_Q \over r_g},
\ee
and express the linear action on the Reissner-Nordstr\"om solution in the following  form: 
\beqa\label{lineargravityRN}
&&S = - M c^2    \int   { 3  \over 2    } \varepsilon \sqrt{  \Big(1    -  {b \over \hat{r}}  
- {c \over \hat{r}^2 }  
+ {f \over \hat{r}^3}   - {e \over \hat{r}^4 } \Big)}   ~ {1  \over \hat{r}^2 } d \hat{r}    dt ~,
\eeqa
where 
\beqa
&& b=   \varepsilon +  \hat{r}_Q^2 { 4  (1 +5 \varepsilon) \over 5  },~~~
c=   \hat{r}_Q^2 {(219 - 480 \varepsilon) \over 45} 
+ \hat{r}_Q^4 {8 (17 - 55 \varepsilon) \over 45 },\nn\\ 
&&~~~
f=  \hat{r}_Q^4 {8  (20 - 33 \varepsilon) \over 9 }, ~~~e= \hat{r}_Q^6 {8 (87 - 125\varepsilon) \over 45 }.\nn
\eeqa
If the charge of the black hole is equal to zero, $ \hat{r}_Q=0$, then the action (\ref{lineargravityRN}) reduces to the expression (\ref{lineargravityS}) on the Schwarzschild solution.  
For  the extremal black hole of the change $r_Q =r_g/2$  the  fourth order polynomial under the root in (\ref{lineargravityRN}) is positive for  $r >   r_g/2 $, is equal to zero at $r =  r_g/2  $ and is negative for $r <  r_g/2 $. The  region which is "locked" for the test particles in this case is defined by $r =  r_g/2  $ and  
prevents the appearance of the naked singularities.  

It is helpful to represent the polynomial under the square root in (\ref{lineargravityRN}) in the form 
\be
\Big(1    -  {b \over \hat{r}}  
- {c \over \hat{r}^2 }  
+ {f \over \hat{r}^3}   - {e \over \hat{r}^4 } \Big)= (1- {\hat{r}_1 \over \hat{r}})...(1-{\hat{r}_4 \over \hat{r}})~,
\ee
where $\hat{r}_i,~i=1,...,4$ are the roots of the fourth order polynomial.
The largest positive real valued root  at which the polynomial turns out to be negative is defined as $\hat{r}_4$. Near that radius one
can approximate the polynomial as 
\be
(1- {\hat{r}_1 \over \hat{r}_4})(1- {\hat{r}_2 \over \hat{r}_4})(1- {\hat{r}_3 \over \hat{r}_4})~(1-{\hat{r}_4 \over \hat{r}}) = \Gamma(\hat{r}_1,...,\hat{r}_4) ~(1-{\hat{r}_4 \over \hat{r}}).
\ee
Thus the action will take the form
\beqa\label{lineargravityRN1}
&&S\approx  - M c^2  \Gamma  \int   { 3  \over 2    } \varepsilon \sqrt{   (1-{\hat{r}_4 \over \hat{r}})}   ~ {1  \over \hat{r}^2 } d \hat{r}    dt 
\eeqa
and it has a similar form with the expression which we analysed in the case of the Schwarzschild  solution (\ref{lineargravityS}). As one can see, the expression under the square root in (\ref{lineargravityRN1}) becomes negative at
\be 
r <  r_4(r_Q,\varepsilon)
\ee
and defines  the region which is unreachable by the test particles.

The Table \ref{tbl:largeN} presents the values of the radius $\hat{r}_4$ at which the polynomial under the root changes its sign from positive to negative and the action becomes complex as a function of the 
charge $\hat{r}_Q$ and the parameter $\varepsilon$.  As it follows from the Table \ref{tbl:largeN} for the extremal black hole, $\hat{r}_Q  =1/2$,  the "locked" region has the radius  $\hat{r}_4 =  1/2 $  and increases with the charge $\hat{r}_Q >1/2$,  preventing the appearance of naked singularities. . At  $\hat{r}_Q =1$ the locked region has the radius $\hat{r}_4 \approx 2.14$. For $\hat{r}_Q  < 1/2$ the locked region is smaller than horizon 
 $\hat{r}_4 < 1/2$.
 \begin{table}[htbp]
   \centering
   \begin{tabular}{@{} lcccrcl @{}}        \toprule
      Charge & Parameter  & Maximal real solution  \\
      $\hat{r}_Q$    & $ \varepsilon$ &  $\hat{r}_4$    \\  
      \midrule
        1   & 0.1 & 2.14   \\
	1/2 & 0.1 & 0.50   \\
        1/4& 0.1 & 0.47   \\
 	1/8 & 0.1 &0.29   \\
	1/16 & 0.1 &0.18    \\
      \bottomrule
   \end{tabular}
   \caption{ The table of  the solutions $\hat{r}_4$ at which the four order polynomial under the root function in (\ref{lineargravityRN})  becomes negative.  The value of $\hat{r}_4$ is measured  in $r_g$ units  (\ref{units}) . For the extremal black hole $\hat{r}_Q  =1/2$  the "locked" region has the radius  $\hat{r}_4 =  1/2 $  and increases with the black hole charge $\hat{r}_Q >1/2$, thus  preventing the appearance of naked singularities. For $\hat{r}_Q  < 1/2$ the locked region is smaller than horizon 
 $\hat{r}_4 < 1/2$, that is $r_4 <  r_g / 2$.}
   \label{tbl:largeN}
\end{table}

 \section{\it Kerr  Solution  }

Let us also consider the Kerr metric  
\be
ds^2 = (1-{r_g r \over \rho^2}) c^2 dt^2 - {\rho^2 \over r^2 - r_g r +a^2} dr^2 -\rho^2 d\theta^2
-(r^2 +a^2 + {r_g r a^2 \over \rho^2 }\sin^2\theta)\sin^2\theta d\phi^2  + 2 {r_g r a \over \rho^2}\sin^2\theta d\phi cdt~,
\ee
where 
\beqa
&g_{00} = 1-{r_g r \over \rho^2},~g_{11} =- {\rho^2 \over r^2 - r_g r +a^2},~g_{22} =-\rho^2,~
g_{33} =-(r^2 +a^2 + {r_g r a^2 \over \rho^2}\sin^2\theta)\sin^2\theta,\nn\\
&g_{03}= g_{30}= {r_g r a \over \rho^2}\sin^2\theta 
\eeqa
and 
$$
a = {J\over Mc },~~~r_g = {2GM \over c^2},~~~~ \rho^2 = r^2 +a^2 \cos^2\theta, ~~~~\sqrt{-g} = \rho^2 \sin\theta.
$$
The nontrivial quadratic curvature invariant is
\beqa
&&I_0={1\over 12} R_{\mu\nu\lambda\rho}  R^{\mu\nu\lambda\rho} = 
{r^2_g(r^2-a^2 \cos^2\theta) [ (r^2 + a^2 \cos^2\theta)^2 -16 a^2 r^2 \cos^2\theta] \over  (r^2 + a^2 \cos^2\theta)^6} 
\eeqa
and it shows that the singularity located at 
$
r=0,~~~\theta = \pi/2
$
is a curvature singularity.
The event horizon is defined by the largest root of the equation 
$r^2 - r_g r +a^2=0$ where the metric component $g_{11}$ diverges:    
\be\label{horiz}
r_{horizon}= {1\over 2}(r_g + \sqrt{r^2_g-4 a^2}).
\ee
For $a > r_g/2$  there are no  real valued  solutions and there is no event horizon. With no event horizons to hide it from the rest of the universe, the black hole ceases to be a black hole and will instead be a naked singularity.
The outer ergosurface is defined by the equation where the purely temporal component $g_{00}$ of the metric changes the sign from positive to negative:
\be
r_{ergosur}= {1\over 2}(r_g + \sqrt{r^2_g-4 a^2 \cos^2\theta}).
\ee
These two critical surfaces are tangent to each other at poles $\theta = 0,\pi$ and they exist only when $a < r_g/2$.  The space between these two surfaces defines the ergosphere. At maximum value of the angular momentum $a = r_g/2$ these surfaces are defined by the equations
\be\label{maxang}
r_{horizon}= {r_g\over 2}~,~~~r_{ergosur}= {r_g\over 2}(1 + \sin \theta).
\ee
Let us now consider the expressions for the curvature polynomials 
$I_1$ and $I_2$ in the case of Kerr solution
\beqa
&&I_1=  
 {r^2_g(r^2- r_g r + a^2 \cos^2\theta) [ r^8 -28 a^2 r^6 \cos^2\theta +70 a^4 r^4 \cos^4\theta - 28 a^6 r^2 \cos^6\theta + a^8  \cos^8\theta] \over  (r^2 + a^2 \cos^2\theta)^9},\nn\\
&&I_2  = {r r^3_g (r^2 - 3 a^2 \cos^2\theta) (r^6 - 33 a^2 r^4 \cos^2\theta 
+27 a^4 r^2 \cos^4\theta - 
   3 a^6 \cos^6\theta) ) \over (r^2 + a^2 \cos^2\theta )^9 }.
\eeqa
It is convenient to introduce the  dimensionless quantities 
\be\label{unitsk}
\hat{r} = {r \over r_g},~~~~~\hat{a}= {a \over r_g},~~~~~\sigma^2= \hat{a}^2 \cos^2\theta
\ee
so that the linear action will takes the form
\beqa\label{lineargravityK}
&&S = - M c^2    \int   { 3  \over 2    } \varepsilon \sqrt{  f(\hat{r}, \hat{a},\varepsilon,\theta) }   ~ {1  \over \hat{r}^2 } d \hat{r}    dt ~,
\eeqa
where
\beqa\label{poly}
&&f(\hat{r}, \hat{a}, \varepsilon,\theta ) = \Big(1 - {\varepsilon \over \hat{r}} - {27   \sigma^2 \over \hat{r}^2} - {8  \sigma^2 \over \hat{r}^3} 
+ { 36  \sigma^2 \varepsilon \over \hat{r}^3} + {42   \sigma^4 \over \hat{r}^4} + {56   \sigma^4 \over \hat{r}^5} 
- { 126   \sigma^4 \varepsilon \over \hat{r}^5} +\nn\\
&&+ {42  \sigma^6 \over \hat{r}^6} - {56   \sigma^6 \over \hat{r}^7} 
+ {84   \sigma^6 \varepsilon \over \hat{r}^7} - {27   \sigma^8 \over \hat{r}^8} + {8   \sigma^8 \over \hat{r}^9} 
- { 9   \sigma^8 \varepsilon \over \hat{r}^9} + { \sigma^{10} \over \hat{r}^{10}} \Big) 
\Big(1+ {   \sigma^2 \over \hat{r}^2} \Big)^{-9} .
\eeqa
The region which is locked for the test particles is defined by the largest real positive root of the polynomial (\ref{poly}) at which the polynomial turns out to be negative. It is  denoted as $\hat{r}_{10}=\hat{r}_{10}(\hat{a}, \theta,\varepsilon)$.
Let us consider the situation with maximal angular momentum $a=r_g/2$ as in (\ref{maxang}).
 The roots $\hat{r}_{10}(1/2, \theta,\varepsilon,)$ can be found numerically for different parameters of the Kerr solutions.  In the maximal angular momentum case some of the values are: 
\beqa
&&a=r_g/2,~~ \theta = 0,\pi ~~~~ r_{horizon}= r_g/2,~~ r_{ergosur}= r_g/2,~~
r_{10} = 2.6 r_g \nn\\
&&a=r_g/2,~~ \theta = \pi/2, ~~r_{horizon}= r_g/2,~~ r_{ergosur}= r_g,~~
r_{10} =   r_g /2 .
\eeqa
Thus the singularity is unreachable by the test particles. In the case of smaller angular momentum the locked region shrinks and is inside the event horizon.

In the sections 2-4 we considered the perturbation of the exact solutions of the classical gravity. It is a difficult task to find the exact solutions of the equations which follow from the action (\ref{lineargravity}) and (\ref{actionlineargravity23}). We were unable to find exact solutions of these equations, but in order to have an idea how these equations look we shall derive the corresponding field equations in  the most simple  case of the linear action in the next section.

\section{\it    Equation of Motion}

As an example let us consider the simplest form of the linear action:  
\beqa\label{lineargravity12}
 S_L = -{M c  \over 8 \pi} \int  \sqrt{
4 R^{~~~;\mu}_{\mu\nu}  R^{\lambda\nu}_{~~~;\lambda}} \sqrt{-g} d^4x,
\eeqa
where the dimension of the integral is the length  and the action measures the "linear 
size" of the universe.  Using the Bianchi identity $2   R^{~~~;\mu}_{\mu\nu} = \partial_{\nu} R$ 
one can represent the action in the equivalent form:
\beqa\label{lineargravity1}
 S_L =
- {M c   \over 8 \pi}  \int    \sqrt{g^{\mu\nu}
\partial_{\mu} R ~\partial_{\nu} R }~\sqrt{-g }  d^4x.
\eeqa
 The corresponding equations of motion can be obtained by variation of the 
action with respect to the metric $\tilde{g}_{\mu\nu} = g_{\mu\nu} + \delta g_{\mu\nu}$
and $\tilde{g}^{\mu\nu} = g^{\mu\nu} - \delta g^{\mu\nu}$. We have
\be
\delta \Gamma^{\lambda}_{\mu\nu}= {1\over 2} g^{\lambda\alpha}
(D_{\mu} \delta g_{\alpha\nu} + 
D_{\nu} \delta g_{\alpha\mu}- D_{\alpha} \delta g_{\mu\nu})\nn
\ee
then
$
\delta R^{\nu}_{\beta\lambda\alpha}= D_{\lambda} \delta \Gamma^{\nu}_{\beta\alpha} - 
D_{\alpha} \delta \Gamma^{\nu}_{\beta\lambda} \nn
$
and 
\be
\delta R_{\mu\nu}= {1\over 2}(-D^2  \delta g_{\mu\nu} +
D_{\alpha} D_{\mu} \delta g^{\alpha}_{\nu} +D_{\alpha} D_{\nu} \delta g^{\alpha}_{\mu} -
D_{\mu} D_{\nu} \delta g^{\alpha}_{\alpha} )~,\nn
\ee
thus 
\be
\delta R= -R_{\mu\nu}  \delta g^{\mu\nu} +
D_{\mu} D_{\nu} \delta g^{\mu\nu}  -
D^2 \delta g^{\alpha}_{\alpha} \nn
\ee
and 
$
\sqrt{-\tilde{g}} = \sqrt{-g}(1+ \delta g^{\alpha}_{\alpha}/2). 
$
Using the above formulas we can calculate the variation of linear action (\ref{lineargravity1})
\beqa
\delta S_L &\propto & \int d^4x \sqrt{-g} \Big{(}  {-\delta g^{\mu\nu} \partial_{\mu} R ~\partial_{\nu} R+ 
2 g^{\mu\nu} \partial_{\mu} R ~\partial_{\nu} \delta R \over 2 \sqrt{g^{\alpha\beta} \partial_{\alpha} R ~\partial_{\beta} R}} + {1\over 2}\sqrt{g^{\mu\nu} \partial_{\mu} R ~\partial_{\nu} R} ~g_{\mu\nu}   \delta g^{\mu\nu}   \Big{)}= \nn\\
&=&\int d^4x \sqrt{-g}  \delta g^{\mu\nu}   \Big{[}
 \Big{(}  R_{\mu\nu} - D_{\mu} D_{\nu}  + 
g_{\mu\nu} D^2  \Big{)}  ~   D_{\lambda}  \Big{(}   {  \partial^{\lambda} R   \over 
\sqrt{ g^{\alpha\beta} \partial_{\alpha} R ~\partial_{\beta} R}   }   \Big{)} + \nn\\
&+&{1\over 2} \Big{(}   -  {  \partial_{\mu} R   \partial_{\nu} R \over 
\sqrt{ g^{\alpha\beta} \partial_{\alpha} R ~\partial_{\beta} R}   }     +g_{\mu\nu}   
\sqrt{ g^{\alpha\beta} \partial_{\alpha} R ~\partial_{\beta} R}   \Big{)}  \Big{]} =0\nn
\eeqa
and find the field equation 
\beqa
\Lambda_{\mu\nu } = \Big{(}  R_{\mu\nu} -D_{\{\mu} D_{\nu\}}+
g_{\mu\nu} D^2  \Big{)}  ~   D_{\alpha}  \Pi^{\alpha} + {1\over 2} \Big{(}  g_{\mu\nu}~  \Pi^{\alpha}  ~\partial_{\alpha} R - \Pi_{\mu} ~\partial_{\nu} R  ~\Big{)}   =0,
\eeqa
 where we introduced the operator $\Pi_{\mu}$ to represent the equation in a compact form:
 \be
 \Pi_{\mu}  = {\partial_{\mu} R \over \sqrt{ g^{\alpha\beta} \partial_{\alpha} R ~\partial_{\beta} R}  },~~~~~\Pi^2 =1,~~~~~D_{\mu} \Pi^{\nu} ~ \partial_{\nu} R=0. \nn
 \ee
The equivalent form of the equation is  
\beqa
\Big{(}  R_{\mu\nu} - {1\over 2}g_{\mu\nu} R -    D_{\{\mu} D_{\nu\}}  + 
g_{\mu\nu} D^2  \Big{)}  ~   D_{\alpha}  \Pi^{\alpha} + {1\over 2} \Big{(}  g_{\mu\nu}~  D_{\alpha}  (\Pi^{\alpha}  ~ R) - \Pi_{\mu} ~\partial_{\nu} R  ~\Big{)}   =0.~~
\eeqa
Taking the covariant derivative $D^{\mu}$ of the above equation one can get convinced that it 
identically vanishes  as a consequence of the diffeomorphism invariance of the action (\ref{lineargravity12}).  These equations contain the high derivative terms and are highly nonlinear. The classical and quantum gravity  theories with generic higher curvature 
terms and corresponding high derivative equations have been considered in \cite{Stelle:1976gc,Stelle:1977ry,Deser:2002jk} and recently in \cite{Kehagias:2015ata,Alvarez-Gaume:2015rwa}.

If one considers the sum
 \beqa
  && S =-{  c^3\over 16 \pi G  } \int R \sqrt{-g} d^4x 
 - {M c   \over 8 \pi} \int   \sqrt{g^{\mu\nu}
\partial_{\mu} R ~\partial_{\nu} R }~\sqrt{-g }  d^4x,
\eeqa
then the equation will take the form 
\be
R_{\mu\nu} - {1\over 2}g_{\mu\nu} R + {2 G  M   \over c^2  } \Lambda_{\mu\nu}=
0
\ee
 and because $2 G  M  / c^2   = r_g$ we shall have 
\be
R_{\mu\nu} - {1\over 2}g_{\mu\nu} R +r_g \Lambda_{\mu\nu}=0.
\ee 
The equations are highly  nonlinear and we were unable to find the exact 
solutions yet. It is a challenging problem and 
in future work we will analyse the field equations which follow from the action (\ref{actionlineargravity23}).
In a forthcoming publication we shall consider a perturbative solution of these equations imposing spherical symmetry \cite{savvidy1}.  In this paper we have only taken the first steps to describe the phenomena  which are caused by the additional linear term in the gravitational action.

In conclusion I  would like to thank Jan Ambjorn for invitation  and
kind hospitality in the Niels Bohr Institute, where part of the work was done. 
I  would like to thank Alex Kehagias for references and Kyriakos Papadodimas for useful 
remarks. The author acknowledges support by the ERC-Advance Grant 291092,  "Exploring the Quantum Universe" (EQU).  

\section{\it Appendix}

The general form of the linear action has the form: 
\beqa\label{generalactionlineargravity}
 && S_L = 
-   m_P c  \int {3  \over 8 \pi}\sqrt{\sum^{3}_{1}  \eta_i K_i+\sum^{4}_{1}  \chi_i J_i+  \sum^{9}_{1}  \gamma_i I_i}~\sqrt{-g }  d^4x ~,\nn
\eeqa
where the curvature invariants have the form 
\beqa
&&I_0={1\over 12} R_{\mu\nu\lambda\rho}  R^{\mu\nu\lambda\rho},    \nn\\
&& I_1= -{1\over 180}R_{\mu\nu\lambda\rho;\sigma} R^{\mu\nu\lambda\rho;\sigma} ~,\nn\\
&&I_2=+{1\over 36} R_{\mu\nu\lambda\rho}  \Box  R^{\mu\nu\lambda\rho}~,\nn\\
&&I_3= -{1\over 72} \Box (R_{\mu\nu\lambda\rho} R^{\mu\nu\lambda\rho} )=
5 I_1 -I_2 ~, \nn\\
&&I_4= -{1\over 90} R_{\mu\nu\lambda\rho;\alpha} R^{\alpha\nu\lambda\rho;\mu} =   I_1 ~,\nn\\
&&I_5=  - {1\over 18}  (R^{\alpha\nu\lambda\rho}  R^{\mu}_{~~\nu\lambda\rho})_{;\mu;\alpha}   =  5 I_1 -I_2~, \nn\\
&&I_6=  -{1\over 18}  (R^{\alpha\nu\lambda\rho}  R^{\mu}_{~~\nu\lambda\rho})_{;\alpha;\mu}  = 5 I_1 -I_2~, \nn\\
&&I_7=  {1\over 18} R^{\alpha\nu\lambda\rho} R^{\mu}_{~\nu\lambda\rho;\alpha;\mu}  =  I_2 ~,\nn\\
&&I_8= R^{\mu}_{~\nu\lambda\rho;\mu} R^{\sigma\nu\lambda\rho}_{~~~~~;\sigma}~,  \nn\\
&&I_9=  R^{\alpha\nu\lambda\rho} R^{\mu}_{~\nu\lambda\rho;\mu;\alpha} ~,\nn\\
&&J_0=  R_{\mu\nu }  R^{\mu\nu } ~,~~J_1= R_{\mu \nu;\lambda} R^{\mu\nu;\lambda }~,~~J_2=  R^{\mu\nu }   \Box R_{\mu\nu }~,~~J_3=  \Box (R^{\mu\nu }    R_{\mu\nu })~,~~J_4=   R_{\mu \sigma}^{~~~;\mu} R^{\nu\sigma }_{~~~;\nu} \nn\\
&& K_0=  R^2~,~~
 K_1=  R_{;\mu}  R^{;\mu}~,~~
 K_2=  R    \Box R~,~~
 K_3=  \Box  R^2~  . 
\eeqa
The $\eta_i , \chi_i $ and $ \gamma_i$ are free parameters. Some of the invariants can be expressed through others using Bianchi identities.

\vfill
\end{document}